\documentclass{aastex63}
\usepackage[utf8]{inputenc}
\usepackage{lineno}

\newcommand{\rp}{\ensuremath{\right)}}
\newcommand{\lp}{\ensuremath{\left(}}
\newcommand{\rc}{\ensuremath{\right]}}
\newcommand{\lc}{\ensuremath{\left[}}
\newcommand{\rb}{\ensuremath{\right\rbrace }}
\newcommand{\lb}{\ensuremath{\left\lbrace}}

\submitjournal{ApJ}

\shortauthors{Labonté \& Merlis}
\graphicspath{{./}{figures/}}

\begin{document}

\title{Sensitivity of the Atmospheric Water Cycle within the Habitable Zone \\
of a Tidally-Locked, Earth-like Exoplanet}


\correspondingauthor{Marie-Pier Labonté}
\email{marie-pier.labonte@mail.mcgill.ca}

\author{Marie-Pier Labonté}
\affiliation{Department of Atmospheric and Oceanic Sciences, McGill University, Montreal, QC, Canada H3A 2K6}

\author{Timothy M. Merlis}
\affiliation{Department of Atmospheric and Oceanic Sciences, McGill University, Montreal, QC, Canada H3A 2K6}

\begin{abstract}

Synchronously orbiting, tidally-locked exoplanets with a dayside facing their star and a permanently dark nightside orbiting dim stars are prime candidates for habitability. Simulations of these planets often show the potential to maintain an Earth-like climate with a complete hydrological cycle. Here, we examine the sensitivity of the atmospheric water cycle to changes in stellar flux and describe the main underlying mechanisms. In a slowly-rotating, tidally-locked Earth-like atmospheric model, the response to a small (about 10\%) increase in stellar irradiance from a habitable-zone control simulation is examined. The water cycle is enhanced in response to the increased stellar irradiance. While the evaporation increase behaves similarly to the stellar radiation increase, the day-to-night energy transport by the mean circulation is critical to the planet’s precipitation changes. Increased efficiency of the energy transport in a warmer climate shapes the substellar precipitation increase. On the nightside, precipitation changes are weak as a result of the large cancellation between the increased energy transport and the increased longwave emission. The day-to-night energy transport efficiency is sensitive to the variation of the atmosphere’s vertical stratification. Due to weak temperature gradients in upper troposphere and a moist adiabat maintained in the substellar region, variations in the substellar surface temperature and specific humidity govern the increase of the planet's stratification with warming. This suggests a scaling of nightside's precipitation based on the substellar surface thermodynamic changes, a sensitivity that holds over a wider range of stellar irradiance changes.

\end{abstract}



\section{Introduction} \label{sec:intro}
Earth-like planets of nearby stars may be found in a tidally locked configuration because it is a common condition for terrestrial planets orbiting red dwarfs, the most common type of star in our universe (\cite{kopparapu13}; \cite{gaidos16}; \cite{barnes17}). The promising observations of exoplanets by future space telescopes makes it exciting to predict and characterize the various potential climate states this specific type of planet could maintain within the habitable zone (\cite{noda17}; \cite{wolf17b}; \cite{wolf17}). 

As we know from Earth, the stellar radiation's spatial pattern drives the planet's atmospheric circulation and, thus, controls the planet's climate. This is directly tied to the Earth's water cycle, which is sensitive to the atmospheric water vapor content and water vapor transport around the globe, and establishes the large-scale positions of arid and wet regions on Earth. The anthropogenic increase in carbon dioxide concentration and the ensuing global warming modifies the Earth's atmospheric circulation and increases the overall moisture content of the atmosphere. Climate change projections generally feature a wetter deep tropics ($\approx 0^\circ\text{-}10^\circ$ latitude, partly due to larger convergence of water vapor) and a drier subtropics ($\approx 10^\circ\text{-}30^\circ$ latitude,  partly due to larger divergence of water vapor) in response to warming. While there is a large body of research on hydrological cycle changes of Earth's climate, the sensitivity of the hydrological cycle of Earth-like exoplanets is less well understood. We expect tidally-locked Earth-like exoplanets' water cycle to be sensitive to the incoming stellar flux, which varies widely across the habitable zone. The climate variations between exoplanets receiving different stellar fluxes could be analogous to the simulated climate response on Earth under global warming.

The use of general circulation models (GCM) has been crucial to grasp the complexity of the climate dynamics on tidally-locked planets. Their permanent faces would be either too warm (dayside) or too cold (nightside) if they were in a radiative equilibrium balance. However, GCMs have shown that an Earth-like climate can be sustained on synchronously rotating planets because atmospheric flows transport energy from day to nightside; this implies that an atmosphere can be maintained rather than condensing on the nightside, and dayside surface temperatures are suitable for surface liquid water (\cite{joshi97}; \cite{joshi03}; \cite{pierrehumbert10}; \cite{wordsworth15}; \cite{wolf17}). The surface temperature and precipitation spatial structures are controlled by the atmospheric circulation. The mean circulation redistributes the dayside's surplus energy to the nightside which, in addition to the presence of reflective dayside clouds and the dry nightside's thermal emission, helps reduce the expected surface temperature extremes (\cite{joshi03}; \cite{merlisschneider10}; \cite{yangabbot14}). The mean circulation redistributes the atmospheric water vapor between dry and wet regions, affecting the precipitation field (\cite{merlisschneider10}). The atmospheric transport of water is important as it modifies a planet's surface climate and, thus, habitability.

Some novel climate states in the water distribution have been explored on tidally-locked planets (with colder climates): all surface water could be trapped in ice caps at high latitudes or over the whole nightside (\citet{leconte13}; \cite{menou13}; \cite{yang14}) leaving a dry dayside, and snowball state towards the habitable zone's outer edge (\cite{checlair17}). For planets closer to the inner edge of the habitable zone, their warmer atmospheres are likely to be richer in water vapor, and therefore  have an active atmospheric water cycle which can affect the runaway greenhouse threshold. However, the energetic constraints on the water distribution (i.e., on the water vapor sources and sinks of surface evaporation and precipitation) on a tidally-locked planet have not been carefully investigated.

In this work, we examine the sensitivity of the atmospheric water cycle to larger stellar fluxes in a tidally locked Earth-like exoplanet configuration of a GCM. Specifically, to evaluate the main energetic constraints on the evaporation and precipitation fields, we tested the climate system response to an increase in stellar radiation, analogous to simulating a planet closer to the inner edge of the habitable zone, using an idealized, ocean-covered atmosphere GCM (section 2).
We show that larger irradiance results in an intensification of the atmospheric water cycle, with moistening in regions where precipitation exceeds evaporation (section 3). This is modulated by an increase in the efficiency of atmospheric energy transport with warming; this larger energy redistribution offsets the local precipitation changes that would occur in a warmer and moister atmosphere with unchanged atmospheric circulation. Then, we highlight the role of the dayside's substellar region thermodynamic fields at the surface (i.e., temperature and moisture content) in regulating the whole planet's energy redistribution (section 4), which constrains changes in the nightside's precipitation. The understanding is developed for small perturbation ($\approx 10\%$) and is then examined over a wide range ($\approx50\%$) of stellar irradiance (section 5). We discuss our findings and conclude in section 6. 

\section{The model and simulation results} \label{sec:metho}
\subsection{Idealized General Circulation Model} \label{sec:model}
We perform simulations using an idealized atmospheric GCM, a 3D hydrodynamic model with an interactive water cycle, as described in \cite{frierson06}, using the T42 truncation in the spectral dynamical core ($\sim 2.8^\circ$ horizontal resolution) with 30 vertical levels. The model has the quasi-equilibrium convective scheme described in \cite{frierson07} and a comprehensive radiative transfer scheme to include water vapor's interaction with radiation, a similar setup as \cite{merlis13}. This radiative transfer scheme calculates the longwave absorption and emission of water vapor, carbon dioxide, methane, and ozone, as well as the shortwave absorption by water vapor, carbon dioxide, and ozone (\cite{anderson04}). We use the model's clear-sky setting, implying that condensed phases of water are assumed to immediately return to the surface (no absorption or scattering by clouds).

The GCM is an idealized aquaplanet which has a slab ocean surface, without ocean heat transport and sea ice. The surface albedo is set to 0.38, which is higher than that of water, but accounts for the omitted cloud effects on the planetary albedo. Focusing on an Earth-like exoplanet, we use: a $6371 \, \mathrm{km}$ planetary radius, an atmosphere with a similar gas composition (nitrogen dominated) without any $\text{CH}_4$, an Earth-like vertical $\text{O}_3$ concentration (between 0.004 in the lower troposphere to 20ppm in the stratosphere), 300ppm $\text{CO}_2$ concentration, and a total atmospheric mass that results in a 1000hPa (1 bar) surface pressure. The main differences from Earth simulations are 1) the imposed tidally-locked incoming stellar radiation; the planet's rotation rate is set to be equal to its orbital period, so the stellar flux reaches only one face of the planet, and 2) the 30-day rotation rate, simulating a slowly-rotating planet compared to the Earth, following \cite{yangabbot14}. Both the obliquity and the orbital eccentricity are set to zero.

First, a control climate simulation is performed assuming the planet orbits a G-type star within its habitable zone, with a stellar constant $S_o = 1360 \,\mathrm{W \, m^{-2}}$ (same value as the Earth's solar constant). This choice of stellar spectrum is necessitated by the accuracy of the radiative transfer algorithm, which was not designed for a red dwarf stellar spectrum (\cite{yang16}). The results that follow are not likely to depend on the radiative transfer details because of the dominant role of atmospheric thermodynamic changes described in what follows. Then, to examine the climate response of a planet closer to the inner edge of the habitable zone, which implies a planet receiving a greater stellar irradiance, we impose a radiative forcing on the control climate by increasing the stellar constant by $\sim10$\%: the perturbed climate simulation is run using a $1500 \,\mathrm{W \, m^{-2}}$ stellar constant. Additional simulations with stellar constant values of 1200, 1650, and $1800 \,\mathrm{W \, m^{-2}}$ are performed to test the robustness of the results. For all perturbed stellar irradiance simulations, the rotation rate is kept at the control, 30-day value.

\subsection{Control Climate and Its Response to the Perturbation}

\begin{figure}[ht]
    \centering
    \makebox[0pt]{\includegraphics[width=1.05\linewidth]{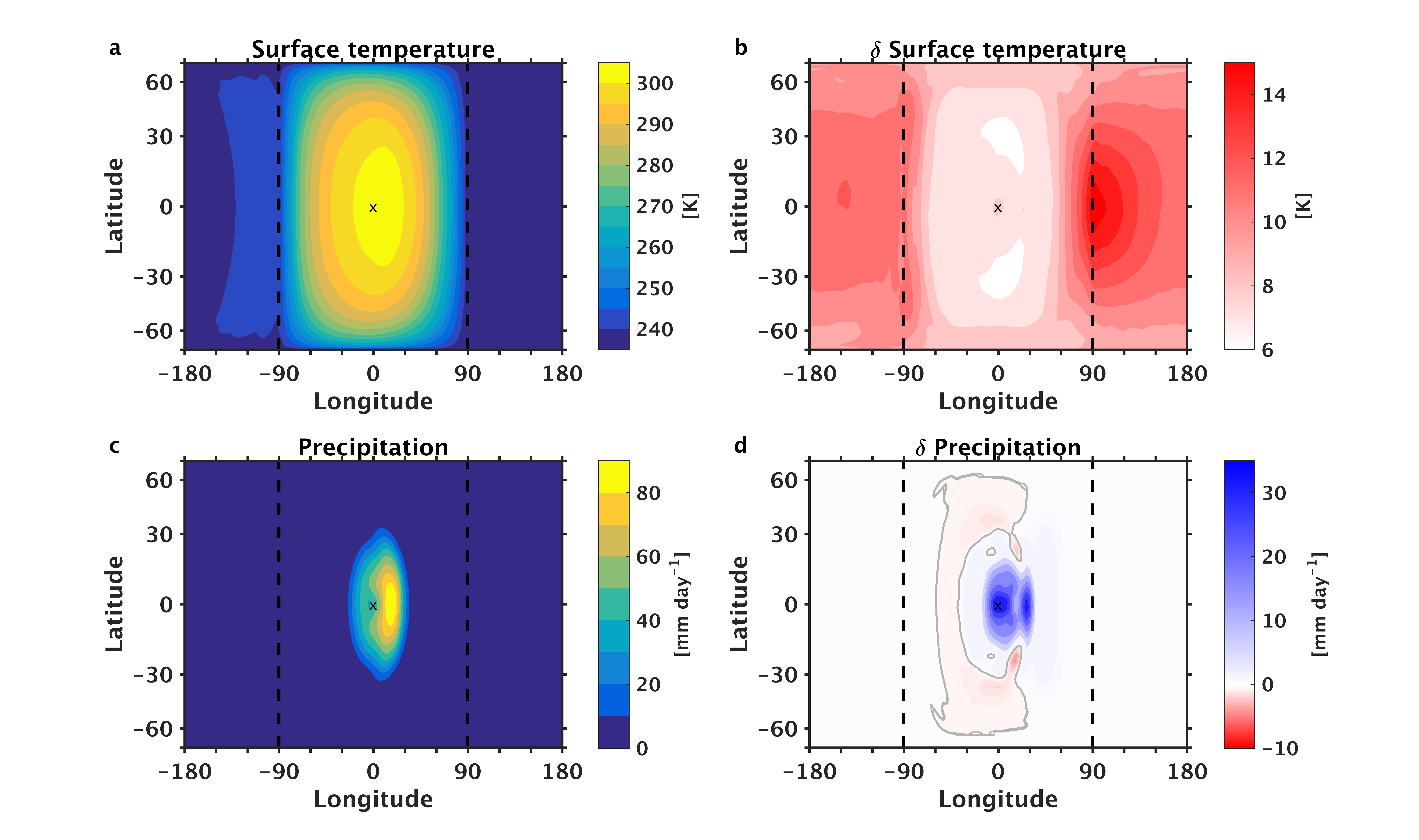}}
    \caption{\small The control simulation's time-mean (a) surface temperature ($5 \,\mathrm{K}$ contour interval) and (c) precipitation rate ($10 \,\mathrm{mm \, day^{-1}}$ contour interval). The response to increased stellar irradiance of the (b) surface temperature ($1 \,\mathrm{K}$ contour interval) and of the (d) precipitation rate (contour intervals of $0.5 \,\mathrm{mm \, day^{-1}}$ for negative values and $5 \,\mathrm{mm \, day^{-1}}$ for positive values, with the zero contour in gray). The x indicates the substellar point (0,0) and the vertical dash-lines represent the terminators between the dayside and the nightside.}
    \label{fig:climate}
\end{figure}

The ``substellar point'' is where stellar rays reach the planet perpendicularly, and is located along the equator at $0^\circ$ longitude (indicated as a cross on Fig. \ref{fig:climate}). The ``substellar region'' refers to the area centered on the substellar point with boundaries at $\sim30^\circ$ in latitude/longitude. In the control climate, the time-mean surface temperature pattern has a north-south symmetry on the dayside, and the maximum temperature of $304 \,\mathrm{K}$ ($\mathbf{31^{\circ}\mathrm{C}}$) is at the equator (Fig. \ref{fig:climate}a). The maximum surface temperature is displaced to the east of the substellar point by advection.  This symmetry breaking results from the planet's non-zero rotation rate, which eventually results in a transition towards the warm surface temperatures distributed in a crescent shape for rapidly-rotating tidally-locked planets (\cite{merlisschneider10}, \cite{showmanpolvani11}). On the nightside, there are small surface temperature gradients and an average temperature of $239 \,\mathrm{K}$ ($\mathbf{-34^{\circ}\mathrm{C}}$). The time-mean precipitation pattern exhibits the same north--south symmetry and east--west (zonal) asymmetry as the temperature on the dayside (Fig. \ref{fig:climate}c). In some regions along Earth's equator, annual-mean precipitation can reach up to $10 \,\mathrm{mm \, day^{-1}}$. This is about 10 times smaller than the $98 \,\mathrm{mm \, day^{-1}}$ maximal rainfall simulated in the substellar region. On the nightside, there is a $\simeq0.01 \,\mathrm{mm \, day^{-1}}$ precipitation rate. This is comparable to the most arid regions on Earth, including the interior of the desert-like Antarctica continent.

In response to the increased stellar irradiance, the nightside's surface temperature increases more than that of the dayside, with approximately two times more warming in some regions (Fig. \ref{fig:climate}b). The largest precipitation increase of $37\,\mathrm{mm \, day^{-1}}$ is located at the substellar point (Fig. \ref{fig:climate}d). In the substellar region, there is a $\approx24$\% precipitation increase and a $\approx7 \,\mathrm{K}$ surface temperature increase; this implies a $\approx3.4$\% precipitation increase per $1 \,\mathrm{K}$ of warming. The Clausius-Clapeyron (CC) relation between the saturation vapor pressure and temperature has a $\approx5.9$\% increase in vapor pressure per $1 \,\mathrm{K}$ temperature increase for the warmest surface temperature of $304 \,\mathrm{K}$. We note that the substellar precipitation rate of increase is lower than the moisture's rate of increase given by the CC relation, a purely thermodynamic estimate of the precipitation increase.

\section{Atmospheric water cycle changes} \label{sec:energybud}
\subsection{Net Precipitation}

\begin{figure}[ht]
    \centering
    \makebox[0pt]{\includegraphics[width=1.05\linewidth]{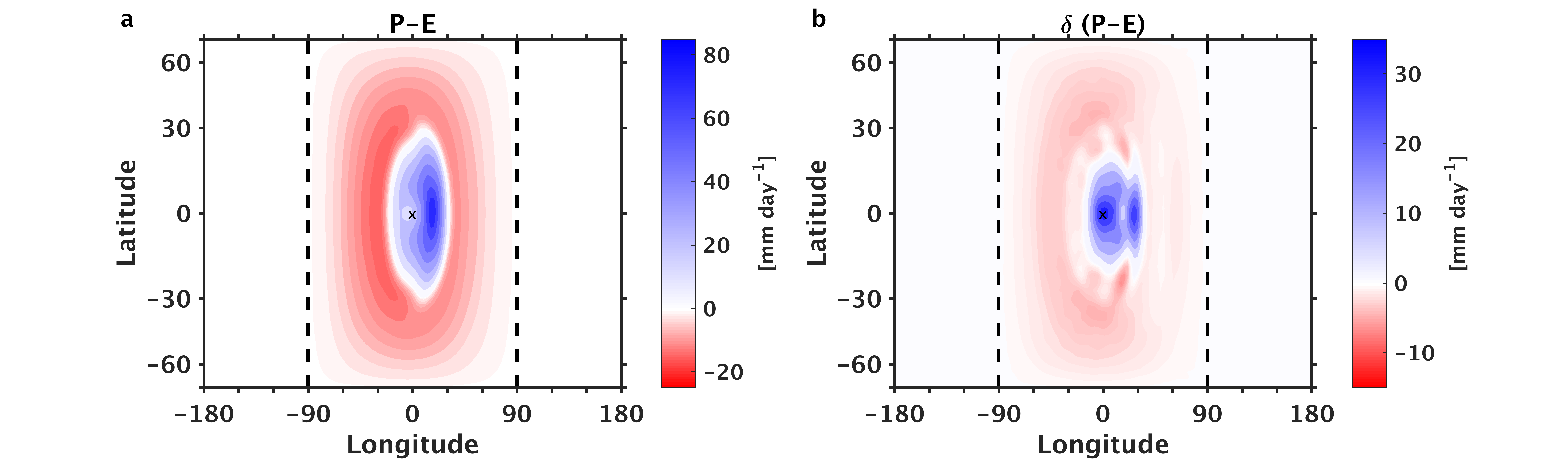}}
    \caption{\small (a) The control simulation net precipitation (precipitation minus evaporation) rate, with exhibiting high precipitation in the substellar convecting region (blue) and net evaporation in the surrounding day-side region (red). (b) The response of the net precipitation rate to increased stellar irradiance, where ``wet regions get wetter, dry regions get drier'' on the day side. The cross indicates the substellar point (0,0) and the vertical dashed lines show the terminator between the dayside and the nightside.}
    \label{fig:netPrecip}
\end{figure}

The atmospheric water cycle's strength is tied to the precipitation ($P$, atmospheric water vapor sink) and the evaporation ($E$, atmospheric water vapor source) rates. Locally, the time-mean net precipitation ($P-E$) indicates the net amount of water vapor converged into an atmospheric column by the atmospheric circulation. Its simulated spatial pattern in the control climate (Fig. \ref{fig:netPrecip}a) displays heavy net precipitation ($P>E$) in the substellar region surrounded by a drier region with low precipitation ($E>P$). These regions of net precipitation and net evaporation are shaped by the atmospheric circulation.

The large-scale atmospheric circulation is driven by the presence of a net energy gain on the dayside and a net energy loss on the nightside. Winds diverge from the substellar region towards the nightside near the tropopause, subside elsewhere and return to the substellar region at the surface, where there is deep convection (analogous to Earth's mean atmospheric circulation in the Tropics). The zonal circulation of a slowly-rotating planet can be idealized as two substellar-to-antistellar-point thermally direct overturning cells (\cite{merlisschneider10}; \cite{leconte13}), analogous to Earth's Hadley or Walker circulations. In the present case (30-day rotation rate), the simulated divergence is stronger in the upper branch of the overturning cell located to the east of the substellar point; it is likely that there is an equatorward momentum flux that accelerates the eastward equatorial flow on the nightside (e.g., Fig. 10 of \cite{showmanpolvani11}). The atmospheric moisture content is replenished in the net evaporation region (analogous to Earth's subtropics near $\mathbf{20^\circ}$ latitude) and the return flow at the surface transports this water vapor towards the substellar region, where it condenses and leads to a large precipitation rate.

The simulated changes of the net precipitation with warming (Fig. \ref{fig:netPrecip}b) show a moistening of the regions with a positive net precipitation rate and a drying of the regions with a positive net evaporation rate: the atmospheric water cycle gets stronger. It is consistent with the ``wet gets wetter, dry gets drier'' description of Earth's hydrological cycle undergoing global warming (\cite{heldsoden06}). This mechanism is tied to an increasing temperature-dependent atmospheric moisture content and the associated change in moisture transport in a warming climate. \cite{heldsoden06} demonstrate that zonally-averaged net precipitation changes are proportional to the lower-tropospheric increase in water vapor: the climatological net precipitation changes simply scale with the CC scaling of $7\,\mathrm{\%/K}$ for global-mean water vapor (assuming a fixed relative humidity). Therefore, the net precipitation pattern is simply enhanced with warming; regions with high precipitation rates tend to get wetter, while regions with evaporation rate exceeding precipitation tend to get drier. The increase in stellar irradiance seems to generate similar changes in the net precipitation pattern on the dayside of the present tidally-locked Earth-like planet.
Next, we show separate analyses of the evaporation and the precipitation changes using the surface and the atmospheric energy budgets, respectively, in the following sections.
\subsection{Surface Energy Budget}

\begin{figure}[ht]
    \centering
    \makebox[0pt]{\includegraphics[width=1.05\linewidth]{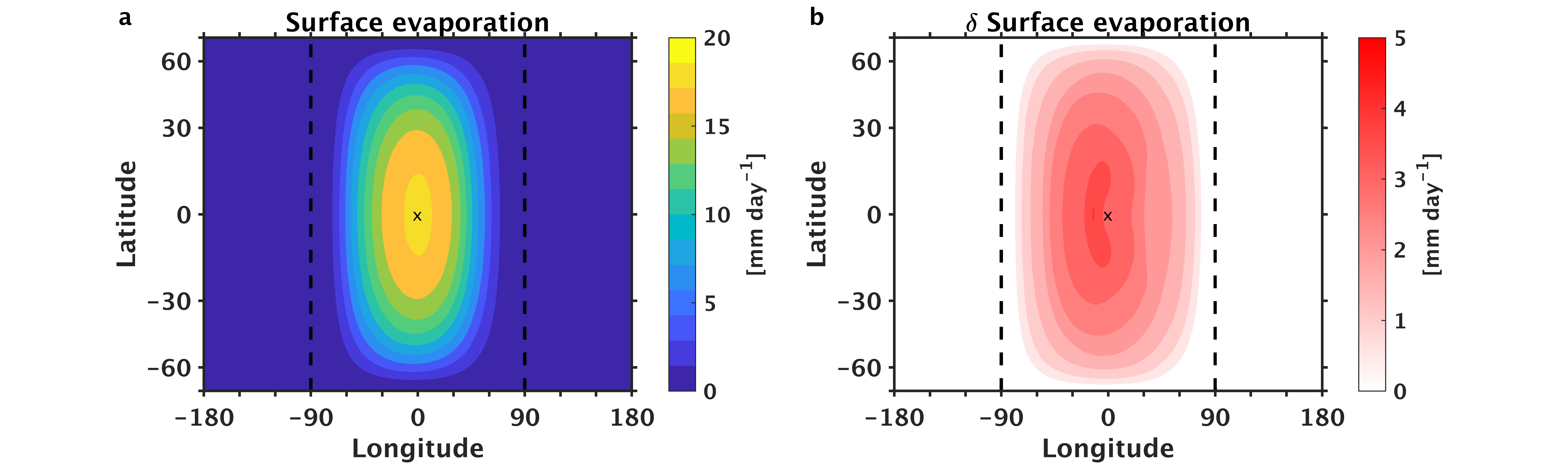}}
    \caption{\small (a) The control simulation surface evaporation rate, which exhibits a similar pattern as the incoming stellar radiation. (b) The response of the surface evaporation rate to increased stellar irradiance. The cross indicates the substellar point (0,0) and the vertical dashed lines show the terminator between the dayside and the nightside.}
    \label{fig:sfcEvap}
\end{figure}

Given the planet surface covered in oceans and a warm-enough climate, the spatial pattern of surface evaporation rate is likely to resemble the incoming stellar radiation, as suggested by \cite{merlisschneider10}, as shown on Fig. \ref{fig:sfcEvap}a. Keeping in mind the absence of ocean energy transport in the present GCM, the surface energy budget in steady-state conditions $R_{sfc} = LE + SH$ is a balance between $R_{sfc}$, the net radiative heating at the surface (the difference between net absorbed stellar radiation $SW$ and longwave emission $LW$ at the surface, $SW^{net}_{sfc}-LW^{net}_{sfc}$), and $LE + SH$, the turbulent latent and sensible energy fluxes cooling the surface. The surface evaporation $LE$ largely balances the net radiative heating over oceans on Earth (about 80\% globally (\cite{hartmann})).

In the control climate, the dayside surface energy budget is composed of the following terms (averaged over 30$^\circ$N-30$^\circ$S): $LE = -266.9\,\mathrm{W\,m^{-2}}$, $LW^{net}_{sfc} = -84.9\,\mathrm{W\,m^{-2}}$, $SH = -36.5 \,\mathrm{W\,m^{-2}}$, $SW^{net}_{sfc} = 386.8\,\mathrm{W\,m^{-2}}$, where a negative sign represent a cooling of the surface, a positive sign a warming of the surface. Consistent with a sun-like star irradiance, the planet surface absorbs $\approx48\%$ of the incoming $SW$ radiation. This number would decrease for a red dwarf star because the planet atmosphere would absorb a larger amount of $SW$ radiation (\citet{shields2019}).

The simulated surface evaporation is not simply balanced by the net absorbed stellar radiation as its magnitude is smaller ($LE\approx 0.69~SW^{net}_{sfc}$) due to significant surface cooling from the other surface fluxes. However, the spatial pattern of surface evaporation is very similar to the one of the net $SW$ at the surface. This is the case since our simulations use clear-sky radiative transfer. Note that, if cloud radiative effects were included in the simulations, we would expect a more complex structure due to the clouds' reflectivity of the incoming shortwave radiation. In the warmer climate, longwave emission and sensible heat flux are smaller at the surface, while the surface evaporation has a greater increase than the increase in net absorbed stellar radiation, so its magnitude becomes $LE\approx 0.79~SW^{net}_{sfc}$. Again, the spatial pattern of surface evaporation is very similar to the one of the net $SW$ at the surface. The warmer the climate, the closer the surface energy budget is to the approximation $SW^{net}_{sfc} \approx LE$, a previously proposed approximation for a limit of global-mean precipitation by \cite{ogorman08}.

The spatial pattern of evaporation changes (Fig. \ref{fig:sfcEvap}b) explain the drying of the desert-like ring (see Fig. \ref{fig:netPrecip}b, drying region surrounding the substellar region); the large evaporation increase overcome the small precipitation changes ($\delta E > \delta P$). The largest evaporation increase is located in the substellar region, however the area is moistening as the larger precipitation dominates the evaporation changes ($\delta P > \delta E$). On the nightside, the control simulation surface evaporation is near zero and weakly negative. This results from the absence of shortwave radiation and the small magnitude of surface longwave emission and sensible heat flux. The nightside’s sensible heat flux is negative (it acts as a surface warming process) because of the temperature inversion present in the lower troposphere on the nightside. However, this temperature inversion does not affect the sign of the net surface longwave emission because the nightside’s atmosphere is dry and, hence, optically thin.  The longwave emission level of the downward component of the surface radiation is located at a high altitude, near the tropopause, where the temperature is colder than at the surface. This results in a positive net surface longwave emission on the nightside (i.e., a surface cooling process). Modest surface evaporation changes are simulated in response to the increased stellar flux due to a modest decrease in sensible heat flux and in surface longwave emission.
\subsection{Atmospheric Energy Budget}

The intensity and the spatial pattern of precipitation rate is tied to the atmospheric column's energy budget. In the atmosphere, the total energy is a sum of kinetic energy, internal energy ($c_pT$), potential energy ($gz$), and latent energy ($Lq$). The sum of the internal and potential energy ($s = c_pT + gz$) is the definition of the dry static energy (DSE) $s$. The rate of change of the DSE ($\text{D}s/\text{D}t$) is approximately balanced by the vertical fluxes of energy due to radiation and sensible energy of the surface ($F_{R+S}$) plus latent heating per unit mass ($Q_{LH}$):

\begin{equation}\label{eq:dseBudget0}
    \frac{\text{D}s}{\text{D}t} = g\partial_p(F_{R+SH}) + Q_{LH}~,
\end{equation}
where the material derivative of $s$ is $\text{D}s/\text{D}t = \partial_ts + \mathbf{u}\cdot\nabla s$, and the kinetic energy transport is neglected (\cite{neelin87}). This dry static energy equation can be used to analyse the regional precipitation response to a radiative forcing (e.g., as used by \cite{mullerogorman11} in the context of Earth's global warming). The vertically integrated, time-mean form of equation \ref{eq:dseBudget0} can be grouped into 1) the DSE flux divergence, 2) the net diabatic cooling ($Q = R - SH$, where $R$ is the net longwave and shortwave radiation in the atmospheric column, evaluated at the surface and top-of-atmosphere, $SH$ is the sensible heat flux at the surface), and 3) the latent heating of precipitation ($LP$). The perturbation of this DSE budget then allows one to relate local surface precipitation changes to $Q$ and the energy transport:

\begin{equation}\label{eq:dseFullBudget}
    L\delta \bar{P} = \delta \lc \bar{Q}\rc + \delta \underbrace{ \lp \lc\mathbf{\bar{u}}\cdot\nabla_h \bar{s}\rc +  \lc\bar{\omega}\partial_p \bar{s}\rc\rp}_{ADV} + \delta \underbrace{\lc\nabla_h\cdot  \lp\overline{\mathbf{u}'s'}\rp\rc}_{TE}~, 
\end{equation}
where $\lc \cdot \rc = \int_0^{p_s} (\cdot)~ dp/g$ indicates the mass-weighted vertical integral over the atmospheric column, $\overline{\lp \cdot \rp}$ is a time-mean, and $\delta$ indicates a perturbation between climates. The middle right-hand side (RHS) term is the mean component of the DSE flux divergence, expressed as the sum of the horizontal and vertical DSE advection by the mean circulation ($ADV$). The last RHS term is the energy transport's transient eddy component ($TE$), evaluated here as a residual. In the figures, we present the terms of the energy budget expressed in units of precipitation rate (i.e., all terms on the RHS are divided by the latent heat of vaporization). Note that in this budget a change in the stellar irradiance only affects $R$ if it is absorbed in the atmosphere, and $SW$ absorption in the atmosphere represents $\approx 21\%$ of the total net incoming stellar irradiance in the control simulation ($\approx 24\%$ in the warmer climate).

\begin{figure}[ht]
    \centering
    \makebox[0pt]{\includegraphics[width=1.05\linewidth]{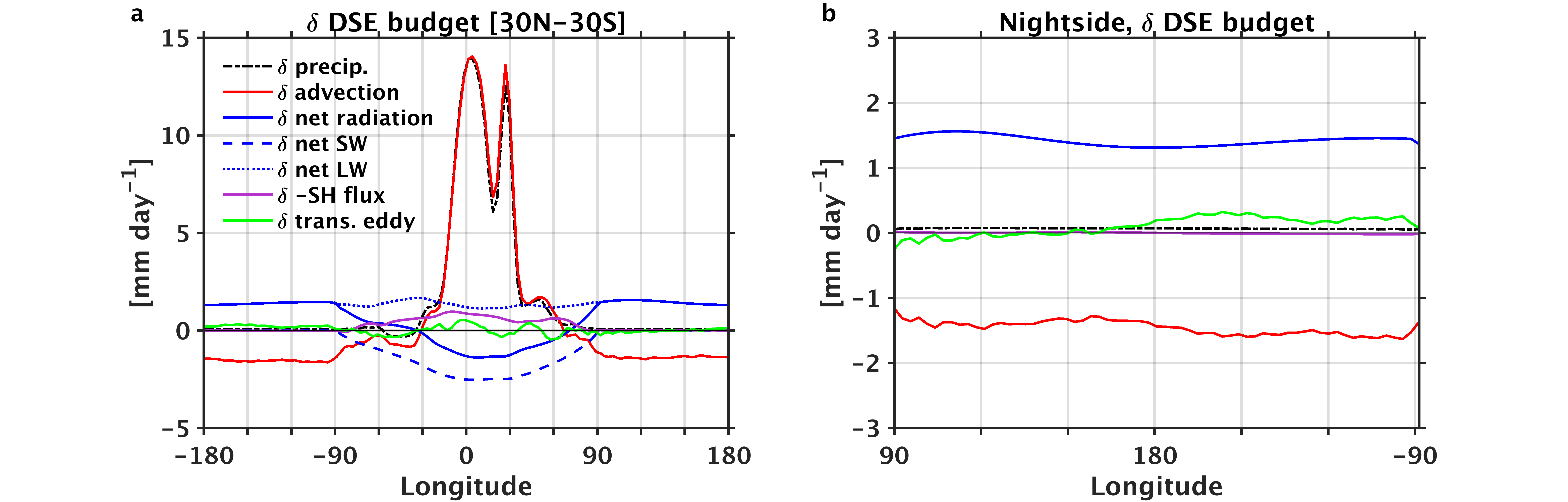}}
    \caption{\small The change in the vertically integrated dry static energy budget (\ref{eq:dseFullBudget}) in response to increased stellar radiation for (a) all longitudes and (b) the nightside. The budget terms include precipitation (black dash-dot line), energy transport (red), the net radiation (blue full line) decomposed into the net absorbed shortwave radiation (blue dash line) and the net longwave emission (blue dash-dot line), the sensible heat flux (purple) and the transient eddies, computed as a residual (green).}
    \label{fig:dseBudget}
\end{figure}

Figure~\ref{fig:dseBudget} shows the perturbation dry static energy budget (eqn.~\ref{eq:dseFullBudget}). The dayside's dominant balance is between the precipitation rate and the divergence of energy transport. The simulated diabatic cooling changes $\delta Q$ are small since $\delta R$ and $\delta SH$ are of opposite signs. The atmospheric column is warmed by a positive net radiation on the dayside, which increases as the incoming irradiance increases (Fig. \ref{fig:dseBudget}a, blue full line). The simulated $\delta SH$ shows that the sensible turbulent flux decreases with warming (purple line), which results from decreased air-sea temperature difference as the climate warms as evaporation becomes the even more dominant turbulent flux. Overall, the increase in stellar flux does not cause large changes in $Q$, because the increase in atmospheric absorption of $SW$ (blue dash line) is partly compensated for by increased $LW$ cooling (blue dash-dot line), preventing larger $\delta R$. There is also a decrease in sensible heat flux which also limits the overall changes in $Q$. Therefore, the precipitation and the energy advection have a similar sensitivity to the increased stellar flux (black dash-dot and red lines). This implies $L\delta P \simeq \delta \lp ADV \rp$: the energy export by the mean circulation is the principal cooling process on the dayside; this greater cooling tendency in a warmer climate is balanced by the warming effect of a larger precipitation rate.
Note that we chose to present results with an average over the latitudes 30$^\circ$N-30$^\circ$S (neglecting high latitudes), consistent with the boundaries of the substellar region that we defined. This is to focus on the convective region. With a 90N-90S average, results show more spatial variations, but overall the same balance remains.

The nightside's energy balance adjusts in a somewhat different manner with warming (Fig. \ref{fig:dseBudget}b); the precipitation is the residual of a dominant balance between net radiative cooling and the energy transport. The simulated diabatic cooling changes $\delta Q$ are relatively large because of the net radiative cooling $\delta R$ increases with warming via increased $LW$ emission, while $\delta SH$ changes little. The simulated $\delta SH$ is near zero due to cold surface temperatures (purple line). The atmospheric column is radiatively cooled by the net atmospheric thermal emission, which increases as the incoming irradiance gets larger (Fig. \ref{fig:dseBudget}b, blue line). As discussed by \cite{yangabbot14}, the nightside is an effective thermal emitter due to the dryness of its atmosphere and the temperature inversion in the lower troposphere. Accordingly, the atmospheric thermal emission is the principal cooling process on the nightside. If the energy budget was similar to the dayside, the precipitation rate would balance the cooling tendency, meaning $L\delta P \simeq \delta\lc R\rc$: a large radiative cooling from the warm troposphere would lead to a substantial precipitation increase at the surface. However, the simulated precipitation rate change on the nightside is very small, with a weak response to the increased irradiance. 
The day-to-night atmospheric energy transport damps the thermal emission cooling effects. In a warmer climate, both the thermal emission and the energy import from the dayside get larger. As shown on Fig. \ref{fig:dseBudget}b, the nightside’s precipitation response to warming (black dash-dot line) is the residual of the opposing cooling tendency of thermal emission (blue line) and warming tendency of the day-to-night heat transport (red line). This implies $L\delta P \simeq \delta \lc R\rc + \delta \lp ADV \rp$: both RHS terms have a similar response to increased irradiance, but of opposite signs, hence precipitation changes remain small.

The atmospheric water cycle intensity is tied to the energy transport efficiency on tidally-locked planets. The latter offsets the local precipitation changes on the dayside, and restricts the amount of water vapor precipitating out on the nightside. In the following subsection, we pinpoint the simplest explanation for the increased energy redistribution with increased irradiance. 
\subsection{Energy Transport Efficiency}

\begin{figure}[ht]
    \centering
    \makebox[0pt]{\includegraphics[width=1.05\linewidth]{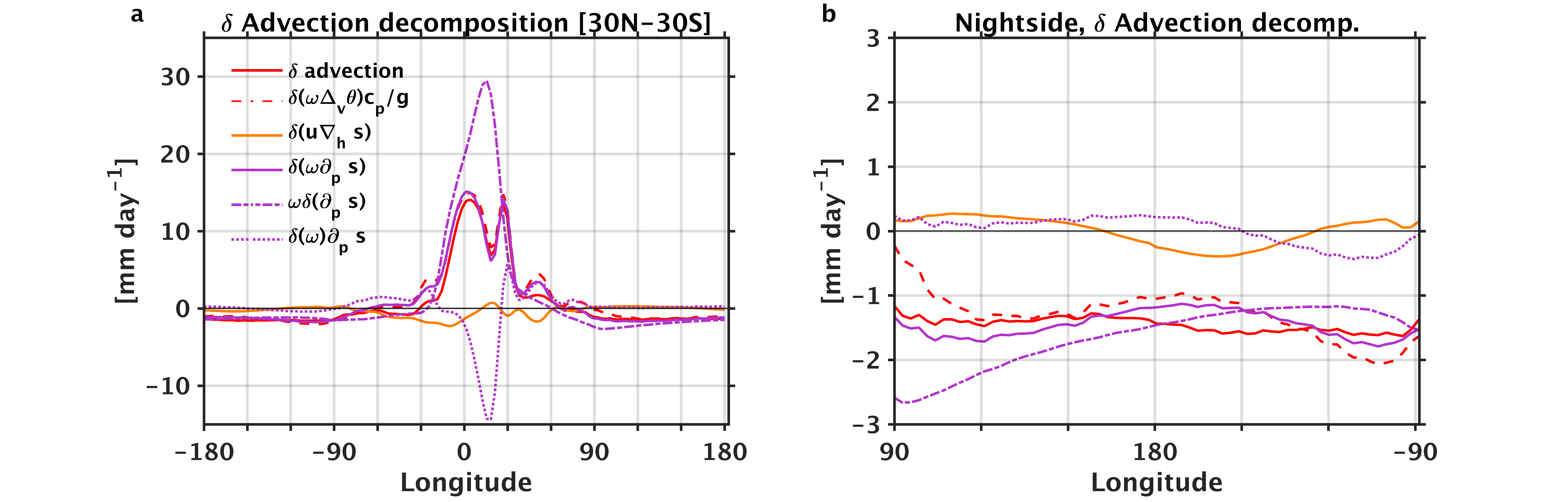}}
    \caption{\small The decomposition of the change in the vertically integrated (surface-to-tropopause) mean circulation's energy transport in response to increased stellar irradiance for (a) all longitudes and (b) the nightside. The total energy transport (red) is decomposed into the horizontal advection (orange), the vertical advection (purple), the mean circulation perturbation (purple dash) and the atmospheric dry stability perturbation (purple dash-dot). }
    \label{fig:advDecomp}
\end{figure}

The larger atmospheric energy transport can be explained by the vertical advection response to warming. The mean circulation energy transport is the sum of the horizontal and the vertical advection in equation \ref{eq:dseFullBudget}. The nightside's vertically integrated DSE horizontal gradient $\nabla_h s$ is small because there is little to no horizontal structure in the surface temperature pattern. Small DSE horizontal gradient is also typically the case in Earth's tropical upper troposphere (\cite{sobel01}) and is expected to hold more globally in the planet's upper troposphere given the slow rotation rate (\cite{pierrehumbert19}). The simulated horizontal advection changes have a weak response to increased stellar flux (Fig. \ref{fig:advDecomp}, orange line). Hence, the simulated vertical advection increase (purple full-line) dominates the increased atmospheric energy transport (red line). This shows that the vertical structure of the dry static energy undergoes a more dramatic change than its horizontal structure. There is an overall increase in the atmosphere's dry static energy; however, the horizontal gradients stay more or less the same while the vertical gradients have a large increase.

\begin{figure}[ht]
    \centering
    \makebox[0pt]{\includegraphics[width=1.0\linewidth]{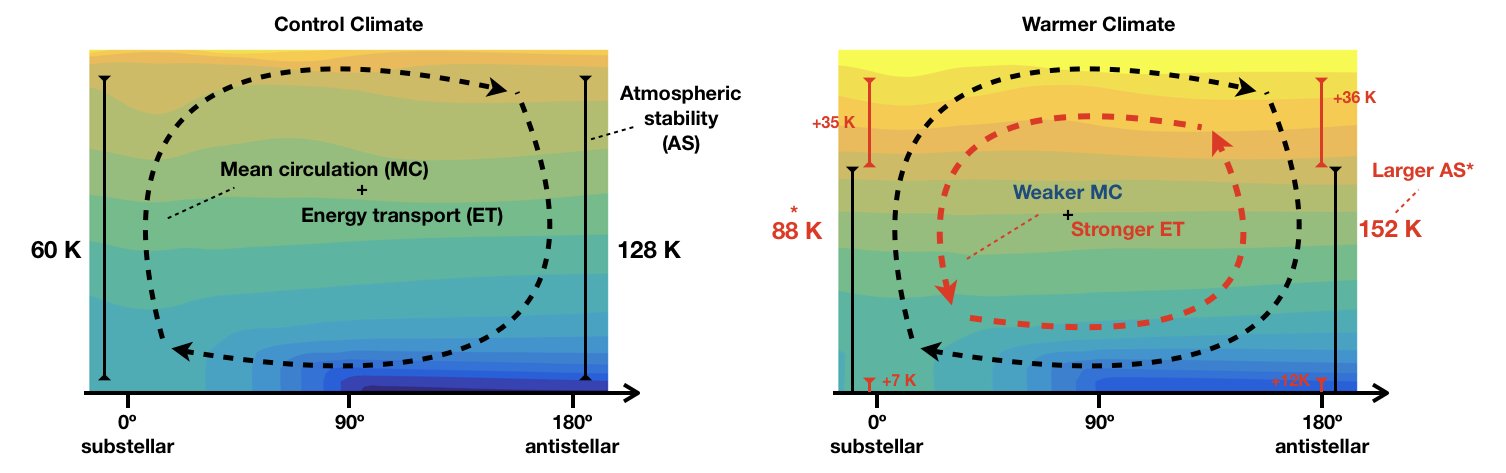}}
    \caption{\small Schematic of the mean circulation mass transport (MC), energy transport (ET), and atmospheric stability (AS) for the (left) control climate and (right) warmer climate. Background contours show the GCM-simulated potential temperature $\theta$ with a 10K contour interval. In the control climate, the atmospheric stability (AS) is $60 \,\mathrm{K}$ in the substellar region and $128 \,\mathrm{K}$ on the nightside. In the warmer climate, the AS increases everywhere: it is $88 \,\mathrm{K}$ in the substellar region and $152 \,\mathrm{K}$ on the nightside. The AS increase ($\delta\Delta_v\theta\approx24\text{-}28\,\mathrm{K}$) occurs because the increase in upper troposphere potential temperature ($\delta\theta^{up}\approx35\text{-}36\,\mathrm{K}$) is greater than the increase at the surface ($\delta\theta^{sfc}\approx7\text{-}12\,\mathrm{K}$).}
    \label{fig:schema}
\end{figure}

The vertical advection of energy $[\bar{\omega}\partial_p \bar{s}]$ is the product of 1) the vertical velocity $\omega$ with 2) the atmospheric vertical stratification $\partial_p \bar{s}$. Figure \ref{fig:schema}a illustrates the relationship between these variables.
 The vertical velocity $\omega$ determines the mean circulation intensity (dashed arrow on Fig. \ref{fig:schema}): upwards movement in the substellar region with weaker downwards winds over the nightside (cf. an atmospheric heat engine perspective was discussed by \cite{koll16}).
The atmospheric vertical stratification is a measure of the resistance to vertical displacements. 
Accordingly, the nightside, region of subsidence has a large stratification compared to the substellar region, which is a region of deep convection. The vertical contrast in DSE $\partial_p\bar{s}$ scales with the vertical difference of potential temperature $\theta = T(p/p_0)^{R/c_p}$ (background contours on Fig. \ref{fig:schema}), where $\theta$ is the temperature an air parcel would have if brought to the surface adiabatically. Hence, it is a tool to compare the temperature/energy of air parcels located at different altitudes. Moreover, the potential temperature definition can be related to other quasi-conserved quantities such as equivalent potential temperature $\theta_e$ as we do in section \ref{sec:atmosstrat}. Consequently, we express the atmospheric vertical stratification as the difference between the potential temperature values at the tropopause (up) and at the surface (sfc), $(\Delta_v\theta)c_p = (\theta^{up} - \theta^{sfc})c_p$. $\theta^{up}$ is evaluated at the pressure level where the divergent winds are the strongest over the substellar region. The vertical advection of potential temperature (Fig. \ref{fig:advDecomp}, red dash) is similar that of DSE (purple), with larger discrepancies on the nightside.

This highlights the two main processes involved in the sensitivity of the total energy transport: the changes in the mean circulation $[\delta\omega\Delta_v\theta]$, and the changes in the atmospheric vertical stratification $[\omega\delta(\Delta_v\theta)]$.
\cite{knutson95} described the possibly counterintuitive result that the response of Earth's tropical atmosphere to carbon dioxide warming consists of a weakened $\omega$ and strengthening $\Delta_v\theta$. An increased stellar flux generates a similar response in our tidally locked configuration.
The simulated mean upward (downward) winds $\omega$ on the dayside (nightside) are reduced with increased irradiance (Fig. \ref{fig:advDecomp}, purple dot). Weaker winds have a warming effect on the dayside since there is a slower export of energy and there is an associated cooling effect on the nightside. While this perturbation is very large in the substellar region, it is weak on the nightside. The vertical stratification increases everywhere with warming, because the potential temperature has a larger increase in the upper troposphere than at the surface and the tropopause rising, as illustrated on Fig. \ref{fig:schema}b. The perturbation induced by the larger atmospheric stratification (Fig. \ref{fig:advDecomp}, purple dash-dot) has a large cooling tendency on the dayside and is partly offset by the slower mean circulation. On the nightside, the warming tendency of the vertical stratification increase scales with the energy transport response to the increased stellar flux.

The atmospheric energy budget can be simplified as follows. On the dayside (ds), the precipitation sensitivity scales with the changes in the mean circulation energy transport, equation \ref{eq:dseBudgetDay}. On the nightside (ns), there is a larger radiative cooling offset by a larger energy transport; the precipitation is the residual of this cancellation, equation \ref{eq:dseBudgetNight}.

\begin{equation}\label{eq:dseBudgetDay}
    \text{dayside}~~~~~L\delta P_{ds} \sim  \delta(\omega_{ds}~\Delta_v\theta_{ds}) \frac{c_p}{g} 
\end{equation}
\begin{equation}\label{eq:dseBudgetNight}
    \text{nightside}~~~~~L\delta P_{ns} \sim \delta LW^{net}_{ns} + \omega_{ns}\delta(\Delta_v\theta_{ns}) \frac{c_p}{g}~,
\end{equation}
where $\delta$ indicates a perturbation between climates, and $\Delta$ is a spatial variation within a climate. There is a simulated decrease of $\omega$, consistent with the atmosphere response to global warming simulated on Earth. However, these changes are weak on the nightside and are overcome by the larger increased stability of the atmosphere. This is why we neglect $\delta\omega_{ns}$ and focus on $\delta(\Delta_v\theta_{ns})$. In short, planets receiving larger irradiance would have a weaker zonal overturning circulation (which has a large effect on the substellar region precipitation) coupled to a more stable dry atmospheric stratification, resulting to a larger energy redistribution from day to nightside, despite the mass circulation slow down. 
The next section is dedicated to the mechanism behind the atmospheric stratification increase and how this allows us to relate the nightside's precipitation changes to the dayside climate.

\section{Theory for the atmospheric vertical stratification} \label{sec:atmosstrat}

A larger incoming irradiance increases the dayside's surface temperature and atmospheric moisture content, causing greater precipitation rate despite a weaker mean circulation. On the nightside, the precipitation response to increased stellar flux is indirect and much weaker in magnitude. The larger irradiance sets the atmospheric stratification increase, which constrains the increasing energy import to the nightside. We first argue that, due to the spatially uniform changes in upper tropospheric $\theta$, the nightside atmospheric stratification has a similar sensitivity to that of the dayside with warming. Then, we argue that the perturbation of substellar's surface thermodynamic fields globally set the changes in the planet's upper tropospheric $\theta$ via convection. The dayside climate, therefore, controls the nightside stratification sensitivity. Hence, we introduce a scaling of nightside's precipitation by accounting for the dayside thermodynamic changes.
\subsection{Vertical Stratification from Dayside to Nightside} \label{subsec:vertstrat}

\begin{figure}[ht]
    \centering
    \makebox[0pt]{\includegraphics[width=1.05\linewidth]{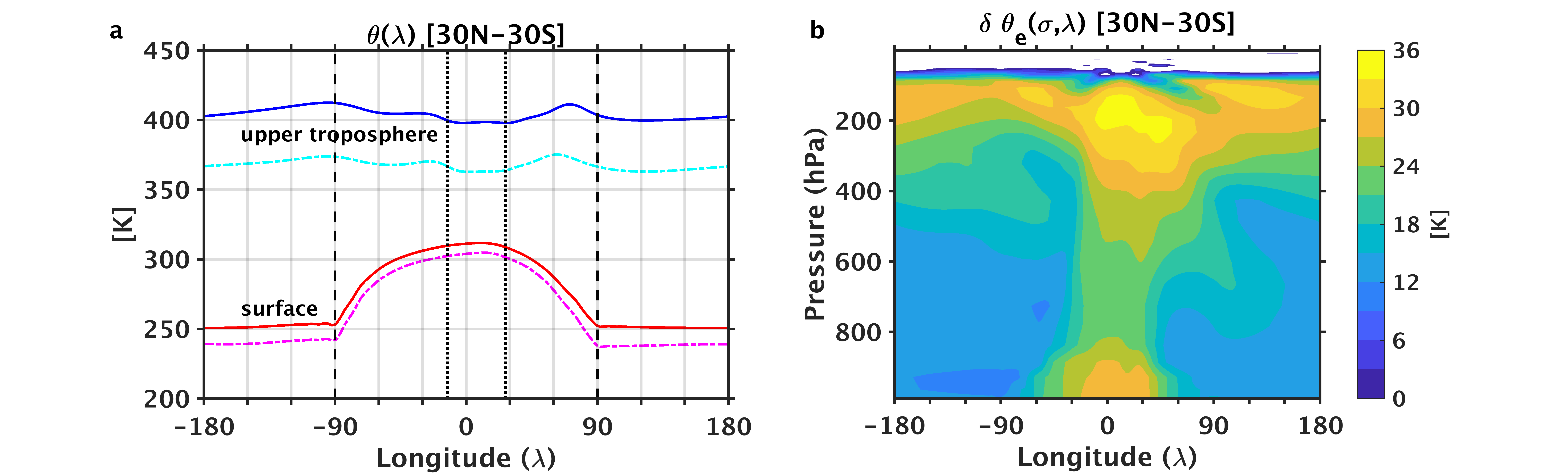}}
    \caption{\small (a) The north--south averaged potential temperature $\theta$ in the control climate (warmer climate) at the surface as the magenta dash-dot line (red full line) and in the upper troposphere as the cyan dash-dot line (blue full line). (b) The equivalent potential temperature $\theta_e$ change between climates (3K contour interval).}
    \label{fig:potTemp}
\end{figure}

Fig. \ref{fig:potTemp}a decomposes the spatial variation of the atmospheric vertical stratification $\Delta_v\theta(\lambda)$ into surface and upper troposphere $\theta$, for both the control climate (cyan/magenta dash-dot lines) and the warmer one (blue/red full lines). First, cyan dash-dot line/blue full line present the upper tropospheric potential temperature $\theta^{up}$ along the tropopause level; potential temperature gradients are weak ($\approx10 \,\mathrm{K}$) over the whole planet. 

In fact, the weak temperature gradient (WTG) approximation is expected to hold in the free troposphere, as the 30-day rotation rate is slow enough for the Coriolis parameter effects to be weak, analogous to what is observed across Earth's Tropics (\cite{sobel01}; \cite{showman13}). WTG is a common feature of tidally-locked exoplanet simulations (\cite{millsabbot13}; \cite{pierrehumbert19}). Second, magenta dash-dot line/red full line present the potential temperature's zonal variation at the surface level $\theta^{sfc}$: the surface temperature steadily decreases from the substellar region to the nightside (Fig. \ref{fig:climate}).

Since there is an approximately uniform potential temperature in upper troposphere, the spatial variation of the vertical stratification can be attributed only to the horizontal surface temperature gradient. \cite{held01b} presented a similar argument through his discussion on the energy transport within the Hadley cell on Earth. He related the atmospheric poleward energy transport in the Tropics to the spatial variation of a moist measure of atmospheric stratification via the spatial variation of surface temperature and specific humidity relative to the convecting region near the equator. The corresponding substellar's convection region (defined as the region with upward mean vertical velocity) is delimited by the vertical black dotted lines around longitude $0^\circ$ on Fig. \ref{fig:potTemp}a. Given the vertical stratification in this region ($\Delta_v \theta_{sub}$), the spatial variation in stratification (a function of longitude) can be expressed as follow:

\begin{equation}\label{eq:stratificationNight}
    \Delta_v\theta_{ns}(\lambda) \approx \Delta_v\theta_{sub} + \Delta_h T_{sfc}(\lambda)~,
\end{equation}
where $\Delta_h T_{sfc}(\lambda) = T_{sub} - T_{ns}(\lambda)$, the substellar region's zonally-averaged surface temperature minus the longitude-dependent surface temperature. In summary, the nightside stratification is the sum of the substellar region's stratification and the surface temperature contrast from day to nightside.

Surface changes $\delta\theta^{sfc}\approx7$-$12\,\mathrm{K}$ (Fig. \ref{fig:potTemp}a, magenta-to-red) are smaller than that of the upper tropospheric changes $\delta\theta^{up}\approx 35$-$36\,\mathrm{K}$ (cyan-to-blue), leading to a vertical stratification increase of $\delta\Delta_v\theta_{sub}\approx 28\,\mathrm{K}$ and $\delta\Delta_v\theta_{ns}\approx 24\,\mathrm{K}$. 
Since surface temperature increases more on the nightside than on the dayside, the nightside vertical stratification undergoes a smaller increase. The horizontal surface temperature contrast change between substellar and nightside is $\delta\Delta_h T_{sfc}(\lambda)\approx -5\,\mathrm{K}$ in the simulations. Neglecting $\delta\Delta_h T_{sfc}(\lambda)$ introduces a modest $\approx15$\% error, hence we can approximate equation \ref{eq:stratificationNight} as $\delta\Delta_v\theta_{ns}(\lambda) \approx \delta\Delta_v\theta_{sub}$. In short, the uniform upper tropospheric changes dominate as the climate warms, hence the nightside atmospheric stratification increase with warming scales with that of the dayside. 
\subsection{Dayside's Surface Thermodynamics Control on Stratification} \label{subsec:sfcthermo}

The dayside's surface temperature and moisture content control the whole planet's atmospheric stratification response by setting the changes in the upper troposphere. The substellar region has large precipitation rates and the deep moist convection in the substellar region maintains a moist adiabatic temperature profile through the depth of the troposphere. Along a moist adiabat, the equivalent potential temperature $\theta_e = \theta \exp\lb L~q/(c_p~T)\rb $ is vertically conserved. The color contours in Fig. \ref{fig:schema} shows the atmospheric stratification in terms of dry air parcels $\theta$. In terms of moist air parcels, the atmospheric stratification would be near zero (constant $\theta_e$) in the substellar's convecting region (where a moist adiabat is maintained) and would be similar to Fig. \ref{fig:schema} in dry regions (upper troposphere and nightside). This is because, unlike $\theta$, the equivalent potential temperature $\theta_e$ takes in account the latent heat released following water vapor condensation in moist adiabatic ascent, and there is a large amount of latent energy $Lq$ in the substellar region's lower troposphere because of its large moisture content.

The region near the tropopause is expected to be dry; the remaining water vapor content in the convective air parcels is approximately negligible ($q_{up}\rightarrow0$) implying $\theta_{e}^{up}\approx\theta^{up}$. Additionally, since $\theta_e$ is a quasi-conserved quantity in convecting regions such as the substellar region, its value at the surface can be expected to be approximately similar to the one in the upper troposphere, $\theta_{e}^{up}\approx\theta_{e}^{sfc}$. Therefore, an equivalence between the moist surface and the dry upper troposphere arises: $\theta_{e}^{sfc}\approx\theta^{up}$. When perturbed, the relationship should remain $\delta\theta_{e}^{sfc}\approx\delta\theta^{up}$. Fig. \ref{fig:potTemp}b shows the $\theta_e$ variation with warming: even though the changes over the substellar region in upper troposphere are greater than that of the surface one, the nightside upper tropospheric changes have the same magnitude as the substellar surface. Moreover, the substellar surface changes $\delta\theta_e^{sfc}\approx27$ to 30K scales with the dry stratification changes $\delta\Delta_v\theta_{sub}\approx28K$.

Therefore, the thermodynamic fields $q_{sfc}$ and $T_{sfc}$ at the substellar region's surface indirectly control the changes in the potential temperature along the tropopause $\delta\theta^{up}$, and thus the whole planet's atmospheric stratification $\delta\Delta_v\theta$. The dependency is mathematically evident as we re-express the substellar stratification, $\Delta_v\theta_{sub} = \theta^{up} - \theta^{sfc}$, within this moist adiabatic framework. To do so, we use the above approximation $\theta^{up} \approx \theta_e^{sfc}$ and the equivalence $\theta^{sfc} = \theta_e^{sfc}~\exp\lb -L~q/(c_p~T)\rb_{sfc} $ to obtain the following expression for the substellar's atmospheric vertical stratification:

\begin{equation}\label{eq:stratificationSub0}
    \Delta_v\theta_{sub} \approx \theta_{e, sub}^{sfc}\lp 1 - \lb e^{- (L~q)/(c_p~T)}\rb_{sfc, sub} \rp
\end{equation}

Applying a Taylor series expansion to the exponential term further simplifies equation \ref{eq:stratificationSub0} to express the substellar vertical stratification in terms of its surface thermodynamic fields:

\begin{equation}\label{eq:stratificationSub}
    \Delta_v\theta_{sub} \approx \lb\theta_e^{sfc}\frac{L~q_{sfc}}{c_p~T_{sfc}}\rb_{sub}~.
\end{equation}

The moisture and temperature at the surface of the substellar region sets the moist adiabat through the whole troposphere. As the stellar flux is increased, whatever changes occur at the surface, the same variation is expected to ensue at upper tropospheric levels (near tropopause). This and the weak upper tropospheric temperature gradients are the key elements of the nightside and dayside interaction.
In other words, we argue that information about the substellar region's surface is sufficient to understand the main features of the whole planet's energetic balance and hydrological cycle.

\subsection{Atmospheric Energy Budget Scaling}

\begin{figure}[ht]
    \centering
    \makebox[0pt]{\includegraphics[width=0.7\linewidth]{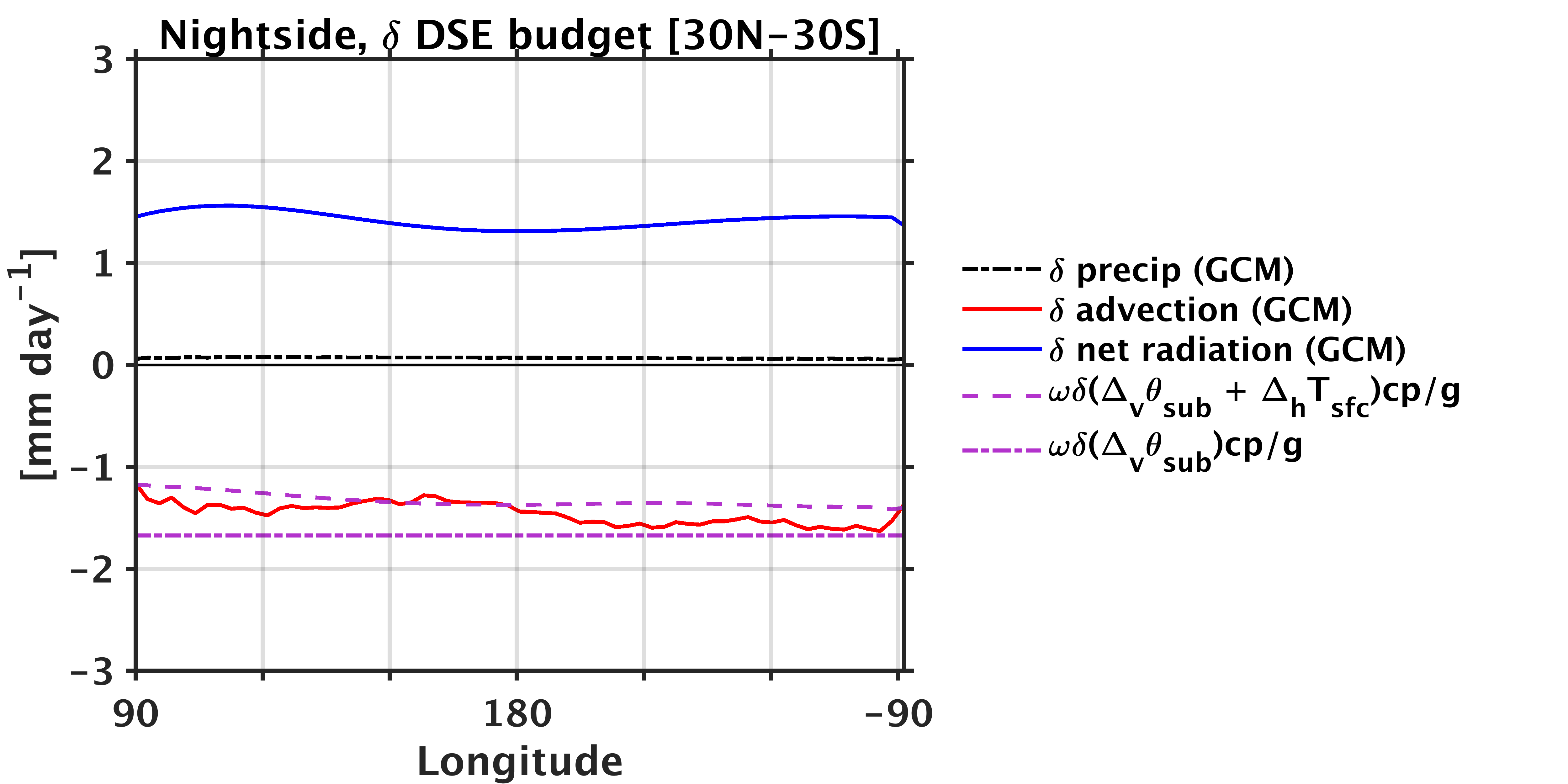}}
    \caption{\small The simplified dry static energy budget on the nightside, equation \ref{eq:dseBudgetNight}, showing the response to the increased stellar irradiance of the GCM simulated precipitation (black dash-dot), net radiative cooling (blue), and total energy transport (red), alongside the approximate energy transport as defined in equation \ref{eq:dseBudgetNightApprox} when the term $\Delta_hT_{sfc}$ is included (purple dash) versus when it is neglected (purple dash-dot).}
    \label{fig:approxEn}
\end{figure}

The atmospheric stratification approximations from the subsections \ref{subsec:vertstrat} and \ref{subsec:sfcthermo}, equations \ref{eq:stratificationNight} and \ref{eq:stratificationSub}, can be included in the approximate nightside atmospheric energy budget, equation \ref{eq:dseBudgetNight}, in order to obtain a new simplified expression:

\begin{equation}\label{eq:dseBudgetNightApprox}
    \text{nightside}~~~~~L\delta P_{ns} \sim \delta LW^{net}_{ns} + \omega_{ns}~\delta\lp \lb\theta_{e}^{sfc} \frac{L~q_{sfc}}{c_p~T_{sfc}}\rb_{sub} + \Delta_h T_{sfc}\rp \frac{c_p}{g}~,
\end{equation}
where $\omega_{ns}$ taken to be a constant over the whole nightside. Fig. \ref{fig:approxEn} presents the new nightside energy budget from equation \ref{eq:dseBudgetNightApprox}, where simulated GCM precipitation, net radiative cooling, and energy transport by the mean circulation are plotted along energy transport approximations. The approximate energy transport is decomposed as follow: the nightside's energy transport sensitivity as defined in equation \ref{eq:dseBudgetNightApprox} (purple dash), and the same nightside's energy transport sensitivity where $\Delta_hT_{sfc}$ is neglected (purple dash-dot). 

The changes in energy transport with increased irradiance largely arise from the substellar's atmospheric stratification sensitivity $\delta\Delta\theta_{sub}$, as expected from Fig. \ref{fig:potTemp}. The changes in the horizontal surface temperature contrast $\delta\Delta_hT_{sfc}$ only adds a correction to the zonal structure; it takes into account the nightside's surface enhanced warming ($\approx2$x that of the dayside), and thus reduces the atmospheric stratification and the amount of heat imported to the nightside. Compared to the simulated GCM total energy transport by the mean circulation (red), there is a slight underestimation of the warming tendency in some regions (purple dash). However, when the nightside's energy import is constrained exclusively by the thermodynamic fields at the substellar's surface (purple dash-dot), it leads to a slight overestimation of the warming tendency in the region with the greatest surface temperature increase.

The nightside radiative cooling has an increase of $\approx2.9\%$ per $1\,\mathrm{K}$ of dayside warming (for a substellar's surface temperature $7\,\mathrm{K}$ increase), while the energy transport, with or without the horizontal surface temperature contrast, has respectively a $\approx2.7\%/\,\mathrm{K}$ and a $\approx3.3\%/\,\mathrm{K}$ increase. The latter is similar to the averaged precipitation response to the climate warming in the substellar region ($\approx3.4\%/\,\mathrm{K}$), consistent with the approximated equation \ref{eq:dseBudgetDay} stating that the dayside precipitation changes are bounded by the energy transport changes with warming. Essentially, we are showing that the nightside precipitation changes are restrained by the dayside climate. A warmer and more humid substellar region leads to a more stable atmosphere and a larger energy export/import on the dayside/nightside. Minimal information about the substellar moisture content and surface temperature allows us to assess the expected energy redistribution to the nightside and the ensuing influence on precipitation rate.

\section{A wider range of stellar irradiance} \label{sec:widerrange}

We have identified the main energetic constraints on the precipitation field and have proposed a scaling for the nightside precipitation based on the thermodynamics of the substellar region surface. To assess the robustness of our arguments, we verify our assumptions in a wider range of climates, with additional simulations using stellar constants of 1200, 1650, and $1800 \,\mathrm{W \, m^{-2}}$.

\begin{figure}[ht]
    \centering
    \makebox[0pt]{\includegraphics[width=1.05\linewidth]{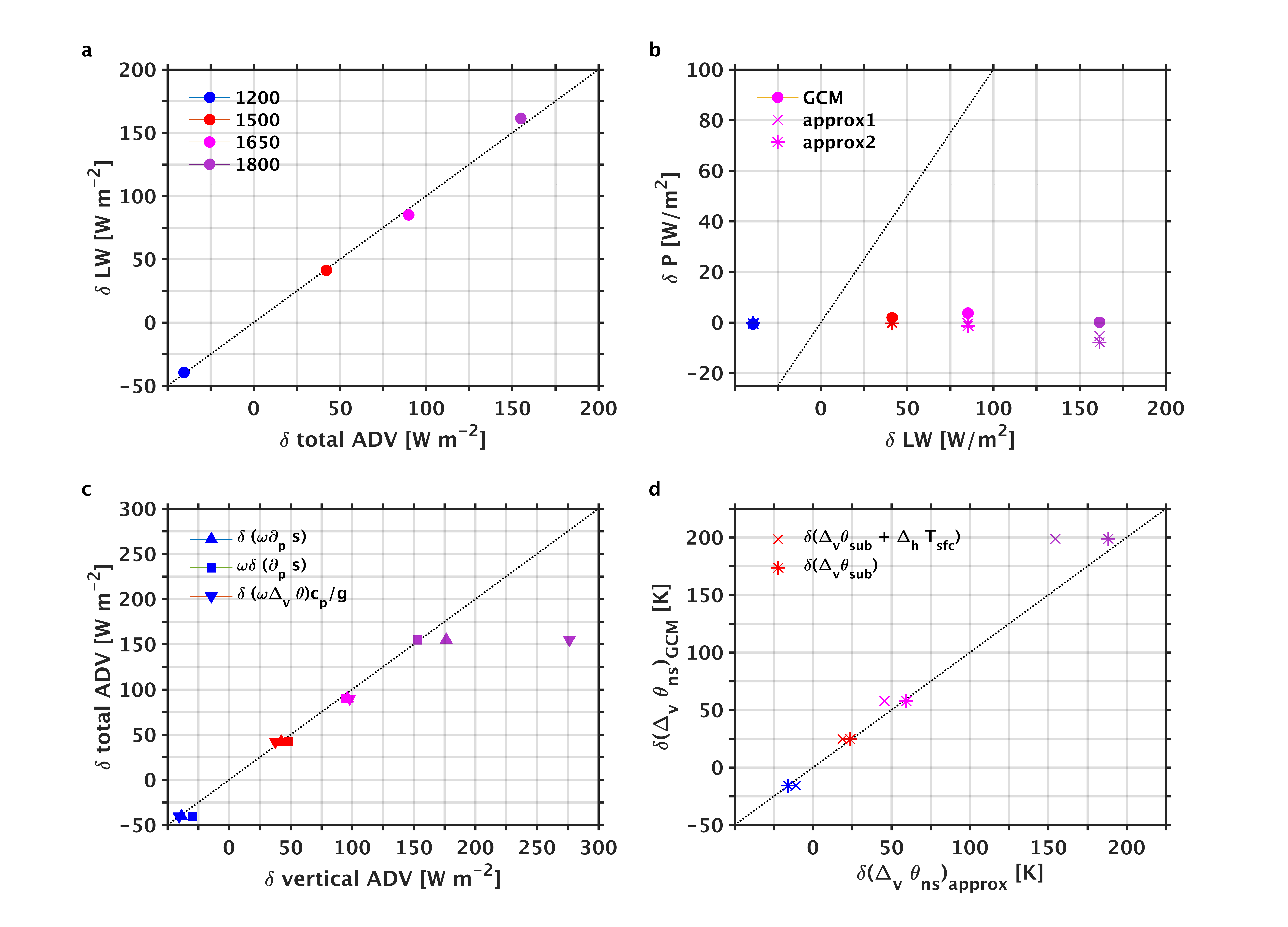}}
    \caption{\small Nightside-averaged energy budget terms relative to the $S_o = 1360\,\mathrm{W \, m^{-2}}$ control climate for a wider range of climates with $S_o = $ 1200, 1500, 1650 and $1800\,\mathrm{W \, m^{-2}}$. (a) The radiative cooling $\delta LW$ vs. the total energy transport by the mean circulation $\delta ADV$. (b) The precipitation $\delta P$ vs. the radiative cooling $\delta LW$, with the GCM-simulated precipitation (circles) and two approximate estimates for precipitation (described below). (c) The total energy transport by the mean circulation $\delta ADV$ vs. the vertical advection, for the total simulated vertical $ADV$, the simulated vertical $ADV$ where vertical velocity is held fixed, and the approximated vertical $ADV$ using potential temperature. (d) The simulated atmospheric stratification using potential temperature vs. the approximate stratification for two approximations. In panels (b) and (d), approximation 1 (`x' symbols) includes the surface temperature contrast $\delta(\Delta_v\theta_{ns}) \approx \delta(\Delta_v\theta_{sub} +\Delta_h T_{sfc})$, while approximation 2 (`*' symbols) only includes the vertical stratification approximation  $\delta(\Delta_v\theta_{ns}) \approx \delta(\Delta_v\theta_{sub})$. Both approximations assume the substellar region dry stratification is given by a moist adiabat: $\delta(\Delta_v\theta_{sub}) \approx \delta\lbrace\theta_{e}~(L~q)/(c_p~T)\rbrace_{sub, sfc}$.}
    \label{fig:widerClim}
\end{figure}

The dominant balance on the nightside remains the same in all climates: the warming effect of energy import by the mean circulation ($\delta ADV$) offsets the cooling effect of longwave emission ($\delta LW$), both increasing at a similar rate with increased irradiance (Fig. \ref{fig:widerClim}a). As discussed in section \ref{sec:energybud}, the nightside precipitation is the residual of this energy balance and should remain low if the changes in radiative cooling and energy transport are of same magnitude. Indeed, the simulated precipitation changes are small through all simulated climates (Fig. \ref{fig:widerClim}b, circles).

The total atmospheric energy transport by the mean circulation ($\delta ADV$) can be well approximated by the vertical advection, the product of the vertical velocity with the atmospheric vertical stratification $\delta(\omega\partial_p s)$ (Fig. \ref{fig:widerClim}c, triangles). The changes in the atmospheric stratification dominate those of the vertical velocity; the weakening of the vertical velocity on the nightside can overall be neglected since it has little impact on the total changes of the vertical advection (Fig. \ref{fig:widerClim}c, squares). This illustrates that, as the total energy transport gets larger with larger stellar flux, the changes in the vertical velocity remain subdominant and the atmosphere gets more stable as the planet warms. As discussed in section \ref{sec:energybud}, we chose to use the potential temperature $\theta$ to describe the nightside stratification and the vertical advection $\delta(\omega\Delta_v\theta)c_p/g$ (Fig. \ref{fig:widerClim}c, reverse triangles). This further approximation fails in the warmest simulated climate ($S_o = 1800 \,\mathrm{W \, m^{-2}}$, i.e. 1800-simulation). The approximate energy import's magnitude is exaggerated because the atmospheric stratification is overestimated when using $\theta$ ($\Delta_v\theta_{ns} = \theta^{up}_{ns}-\theta^{sfc}_{ns}$). The tropopause rises as the climate warms; in the 1800-simulation, our criteria gives the tropopause pressure level to be at $\sim 20$hPa, a very high altitude close to the top-of-atmosphere simulated by the GCM. It could be that the assumption of $\theta$ changes being proportional to the dry static energy changes gets less precise at lower pressures, introducing larger discrepancies in the value of $\theta^{up}_{ns}$. Another potential issue is that the prescribed ozone concentration (at fixed pressure levels) and the high-altitude tropopause implies that some of the stratospheric ozone is in the tropospheric column, increasing the amount of shortwave absorption at these higher tropospheric levels. In summary, both the approximation of thermodynamic variables and model configuration can lead to inaccuracies of the theory in the warmest simulated climate.

In section \ref{sec:atmosstrat}, we introduced further approximations for the atmospheric stratification, which allows us to relate the nightside stratification and nightside energy import to the dayside climate. Fig. \ref{fig:widerClim}d presents the nightside atmospheric stratification approximations in relation to its corresponding GCM value ($\Delta_v\theta_{ns} = \theta^{up}_{ns}-\theta^{sfc}_{ns}$), including a comparison between the approximations where the horizontal surface temperature contrast $\delta\Delta_h T_{sfc}$ is included (crosses) vs neglected (stars). The largest discrepancies arise with the 1800-simulation. Additionally, neglecting the surface temperature gradient results to a better fit with the GCM results, especially in the warmer simulated climates (1650, 1800). 
This tells us that there is a compensating underestimate in the substellar region stratification $\delta(\Delta_v\theta_{sub})$ in these climates. Therefore, including the changes in horizontal temperature contrast $\delta\Delta_h T_{sfc}$ (crosses) underestimates the nightside stratification since the surface temperature gradient between day and nightside decreases more $(\delta\Delta_h T_{sfc}<0)$ as the stellar flux increases.

To get to these approximations, we made two main assumptions. The first assumption was based on the WTG approximation in the upper troposphere. A WTG seems to be maintained roughly in all climates; however, the simulation with a stellar constant of 1800 shows bigger spatial variations (not shown). The second assumption was based on the equivalent potential temperature in the substellar region $\theta_{e, sub}^{sfc}$ being maintained throughout the troposphere. Given that both assumptions hold, we expect the increase of potential temperature on the nightside near the tropopause to be of the same magnitude as the increased surface substellar equivalent potential temperature. The assumptions still partly hold in the 1650-simulation where $\delta\theta_{ns}^{up} \geq \delta\theta_{e, sub}^{sfc}$, while they seems to fail in the 1800-simulation where $\delta\theta_{ns}^{up} \gg \delta\theta_{e, sub}^{sfc}$. For the latter simulation (1800), the discrepancy may arise more because the value of $\theta^{up}_{ns}$ is not a quantitatively adequate measure, rather than a different regime of atmospheric thermodynamics. As a whole, the approximations tend to underestimate (by $\leq20\%$, except the 1800 simulation) the overall nightside atmospheric stratification as the stellar flux gets larger.

The nightside precipitation predictions based on our simplified nightside energy budget in equation \ref{eq:dseBudgetNightApprox} give a decrease of precipitation in the warmer simulated climates (Fig. \ref{fig:widerClim}b, crosses and stars), which is the wrong sign compared to the simulated precipitation changes (circles). Note that the simulated precipitation rates stay around $0.1\,\mathrm{mm\,day^{-1}}$ (about $2.9\,\mathrm{W \, m^{-2}}$), which is very low. Given that the precipitation is a sensitive residual of the nightside's energy budget, modest discrepancies in the approximated energy transport then produce a large discrepancy between the approximated and the simulated precipitation. While these approximations can lead to inaccuracies on the sign of precipitation changes, they encapsulate the important result of the energy import to the nightside cancelling the cooling effects of thermal emission: the absolute precipitation magnitude remains low and the nightside remains a dry hemisphere, even in warm climates. They also highlight the importance of the dayside climate on determining the energy redistribution to the nightside.

\section{Discussion and conclusions}

We examined the atmospheric water cycle sensitivity to the stellar flux by using an idealized atmospheric GCM. Essentially, larger stellar flux leads to an intensification of the atmospheric water cycle. This means larger evaporation and precipitation on the dayside, while precipitation increases modestly on the nightside even though it remains small in magnitude ($\sim 0.1\,\mathrm{mm\,day^{-1}}$).
We explained the precipitation's sensitivity to the stellar irradiance using a simplified dry atmospheric energy budget, which relates the local precipitation changes to the net radiative cooling and the net atmospheric energy transport by the mean circulation. We showed that spatial precipitation changes are not constrained purely by the net radiative cooling because the redistribution of energy between the substellar region and the nightside plays an important role. Interestingly, on the nightside, the rate of energy transport increase with larger irradiance matches the rate of thermal emission increase associated with a warmer surface/atmosphere; this prevents the nightside precipitation from being radiatively constrained and inhibits any large nightside precipitation rates across the habitable zone.

We explained the nightside energy transport sensitivity to larger stellar irradiance by understanding the role of the substellar surface in controlling the redistribution of energy on the planet. The atmospheric energy transport is simply the product of the vertical velocity with the atmospheric vertical stratification. As the climate gets warmer, the zonal circulation weakens while the atmosphere becomes more stable (increased vertical stratification). A larger atmospheric stratification implies that there is a larger energy increase in the upper troposphere than at the surface, which leads to larger energy redistribution between hemispheres: an export of energy upwards cooling the dayside surface and an import of energy downwards warming the nightside surface. The dominant changes in the atmospheric stratification explain why there is a more efficient transport of energy even though the mean circulation is weakened. The dayside and nightside are coupled together through the atmospheric stratification; its spatial variation comes solely from the surface temperature contrast between both hemispheres. In a warmer climate, this contrast decreases, and the stratification has a smaller spatial variation; however, there is an approximately similar vertical stratification increase over both hemispheres because the horizontal gradient changes are subdominant to the vertical gradient changes. The global vertical stratification increase with warming is related to the substellar surface moisture and temperature: the substellar surface thermodynamic fields regulate the upper tropospheric potential temperature variation through the moist adiabat maintained in the region, and these changes in near tropopause potential temperature ultimately control how the atmospheric stratification vary over the nightside.

Our findings suggest that the substellar surface controls the whole planet’s energy redistribution and spatial pattern of precipitation; larger substellar’s surface temperature and moisture content produce large precipitation in the substellar region, and maintain a low nightside precipitation rates across the habitable zone. Given the thermodynamic nature of our arguments, our results may still apply to planets with slightly different atmospheric composition than those of our simulations. Two gases associated with the presence of continents (via silicate rock weathering and volcanism) and complex life forms, such as carbon dioxide (CO$_2$), a greenhouse gas, in the troposphere and ozone (O$_3$) in the stratosphere, could be present at various concentrations or not be present at all in a given Earth-like planet's atmosphere. We performed the same simulations as described in section \ref{sec:model}, but varying the CO$_2$ concentration to 4xCO$_2$ (1200ppm) and 1/4xCO$_2$ (75ppm), and a third set of simulations where the ozone concentration was set to zero. In all three set of simulations, there are no changes in the dominant energy balance and the nightside precipitation scaling holds (not shown). Therefore, it seems fair to assume that the atmospheric water cycle response to stellar flux increase does not depend sensitively on the control climate's atmospheric composition. There may, however, be interesting `non-dilute' effects when the atmospheric water vapor becomes a substantial fraction of the total atmospheric mass (\cite{pierrehumbert16}) toward the inner edge of the habitable zone. 

In the present study, no ocean heat transport, sea-ice, or clouds were included. The ocean heat transport plays a role in warming the nightside, as well as reducing the sea-ice coverage on the nightside and dayside's high latitudes, which reduces the planetary albedo (\cite{hu14}; \cite{yang19}). We also know that planets closer to the inner edge of the habitable zone (stellar flux $>1800 \,\mathrm{W \, m^{-2}}$) have weak ocean heat transport and ice-free surface (\cite{yang19}). Furthermore, by neglecting clouds, we do not introduce any changes in the planetary albedo or longwave emission as the climate warms, even though larger stellar fluxes may lead to a larger cloud coverage in the substellar region, which would stabilize the climate (\cite{yang13}). In the case where the water vapor represents a substantial fraction of the total atmospheric mass (such as the 1800-simulation), the presence of clouds would have only a weak greenhouse effect on the climate (\cite{yang13}). Therefore, we might expect their major impact to be on the planetary albedo (shortwave absorption on the dayside), reducing the dayside surface temperature and day-to-night temperature contrast. The dominant balance of the atmospheric energy budget would most likely  be unchanged. However, the impact of clouds in a more moderate climates is harder to assess because of the uncertainties on the potential cloud feedback, which affects the net radiative cooling of the atmosphere. Hence, given the thermodynamic nature of our arguments concerning the stratification and energy transport, we expect the constraints on the atmospheric water cycle introduced here to hold over a wider range of stellar fluxes ($> 1650 \,\mathrm{W \, m^{-2}}$) when including these additional processes, though the radiative contribution to the dry static energy budget may differ.

The planet rotation rate is a parameter that strongly affects our results. We know from \cite{merlisschneider10} that the atmospheric water cycle (e.g., the distribution of time-mean precipitation) is dependent on the atmospheric dynamics. As the rotation rate gets larger, the Coriolis parameter effects get more important at lower latitudes; this implies that the WTG approximation would no longer be valid, and that our approximations for the energy transport and the nightside precipitation would not hold anymore away from a smaller tropical range of latitudes. The constraints on the atmospheric water cycle on rapidly-rotating planets could have an important dynamic component that was not present here, considering the zonal circulation associated with a superrotating equatorial jet.  

We applied an energetics-based perspective on the atmospheric water cycle to shed light on the expected precipitation distribution for slowly-rotating Earth-like tidally-locked planets across the habitable zone. 
Essentially, the substellar region regulates the whole planet's climate. This warmest and moistest area on tidally-locked planets determines the whole planet's atmosphere stability, energy redistribution, and precipitation rate; we found that minimal information from the substellar region is needed to assess these controls nightside's climate.
Synchronously rotating planets with atmospheres rich in water vapor (as one moves closer to the inner edge of the habitable zone) may have a more extreme climate on the dayside along with a warmer nightside that still remains quite dry. An interesting avenue for future research is to consider the extent to which these broad-scale thermodynamic constraints on the hydrological cycle can be remotely sensed by telescopes.

\acknowledgments
This work was supported by Natural Sciences and Research Council (NSERC), a McGill Space Institute (MSI) fellowship, and a Compute Canada allocation.


\bibliography{refs}
\bibliographystyle{aasjournal}



\end{document}